# IoT BASED SMART ATTENDANCE SYSTEM USING RFID: A SYSTEMATIC LITRATURE REVIEW

**Kashif Ishaq[1*], Samra Bibi[1]**
[1]School of Systems and Technology, University of Management and Technology, 54000, Lahore, Pakistan

**Corresponding Author Email:**
kashif.ishaq@umt.edu.pk

## Abstract:

The use of Radio Frequency Identification (RFID) technology is ubiquitous in a number of businesses and sectors, including retail sales, smart cities, agriculture, and transportation. Additionally, educational institutions have started using RFID to track student attendance, combining this technology with Google Sheets and the Internet of Things (IoT) to build a real-time attendance tracking system. For a thorough examination of the creation of a student attendance system, this paper includes a systematic literature evaluation of 21 major research published on IoT based attendance systems employing RFID. This RFID-based attendance system enables automation, eliminating several problems connected with the manual process, such as time wasting, proxies, and the possibility of losing the attendance sheet, in contrast to the traditional attendance system, which depends on manual signatures. By creating a system that automatically registers students' attendance by merely flashing their student cards at the RFID reader, all the aforementioned difficulties may be successfully addressed. This automated method guarantees attendance monitoring accuracy and dependability while also saving time. This paper's conclusion highlights the significant advantages of implementing an IoT-based attendance system based on RFID technology. The suggested solution provides a trustworthy, efficient, and secure alternative to manual attendance techniques, successfully addressing their shortcomings. This paper offers helpful insights for institutions looking to create a cutting-edge attendance system that increases student involvement and academic achievement by looking at guiding principles, best practices, and the successful resolution of difficulties.

# 1. Introduction:

The "Internet of Things" (IoT) concept has recently acquired popularity in both academics and business. In an Internet of Things (IoT) environment, objects—including people and animals—are given distinctive IDs and the ability to independently exchange data over a network without direct human-to-computer interaction. Every educational institution and other organization facilities uses the manual attendance method, which is time consuming(Shah & Abuzneid, 2019). The time-consuming and unsafe conventional method of taking attendance—calling names or having people write their names on paper—is ineffective. One of the answers to this challenge is an attendance system based on radio frequency identification (RFID). It is possible to utilize this system to record a student's attendance at a school, college, or university. It may also be used to track employee attendance at workplaces (Singhal & Gujral, 2012). RFID is a technology that transfers information from an electronic tag or label affixed to an object through a reader using radio waves in order to identify and track the thing (Shah & Abuzneid, 2019). It is particularly helpful since, based on the tag included, it may separately identify a person or a commodity. It may be completed rapidly, and it often takes a fraction of a second (Lim et al., 2009). EHUOYAN's YHU638 RFID reader was chosen as a result of its low cost and simplicity of use. (Saparkhojayev & Guvercin, 2012). The ESP32 module serves as the system's brain (Singh et al., 2022).

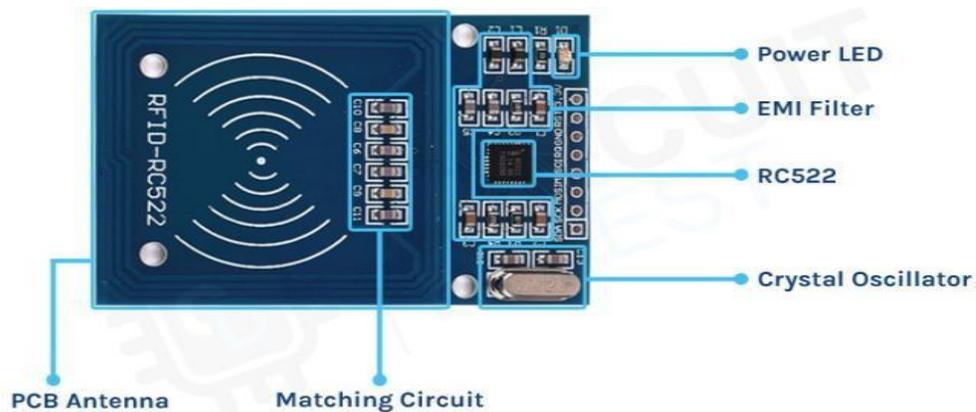

*Figure 1: RFID reader*

The manual attendance method not only takes a lot of time, but it also runs the serious danger of losing important data. The use of such a system has a lot of drawbacks. For instance, lecturers with big classes may find it difficult to acquire the students' manual signatures on the attendance sheet and may find it distracting to educate and gain the students' undivided attention. (Kassim et al., 2012). Monitoring student attendance in the classroom is necessary to improve their academic performance since some students miss class for various reasons and therefore perform poorly on exams (Olanipekun & Boyinbode, 2015). Because it can send accurate data needed as input for software on the attendance system, read data accurately within the reading range, and requires less direct contact between the RFID tag and reader when reading or rewriting data on the RFID tag, RFID technology is superior to other identification technologies. (Ula et al., 2021). Our goal in writing this paper is to use RFID technology to solve this problem. We aim to improve the monitoring of timeliness and successfully manage student and employee absences by implementing a high-quality and well-organized attendance system. A possible method of streamlining attendance monitoring and ensuring a more accurate and seamless record of attendance is the use of RFID.

The essential components of the RFID system are RFID tags, readers, a backend storage system, and a central section that houses all the electrical components. This RFID-based attendance system is particularly user-friendly for commercial usage and contains a storage system that stores the individual identification number of the student or employee (Choudhury, 2017). The system utilizes RFID tags, enabling educational institutions such as schools, colleges, and universities to monitor student attendance. The system automatically uploads attendance data to a Google Sheet notifies parents through SMS when something is happening. If parents use the services, it will deter kids from using the SMS capability to skip class. Our platform also makes it simple to generate reports with just one click. In addition to being affordable, this IoT-enabled RFID reader is also quite portable. In addition, the power supply system is built to automatically convert to battery power in the event of an AC power outage. The device's mobility is ensured by its small size, which makes it simple to carry to schools or other locations. To achieve our goal, we synced both the hardware and software components. Every user's record will be kept up to date and will have a unique RFID tag.

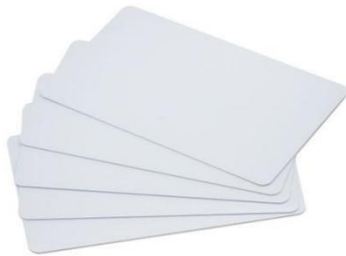
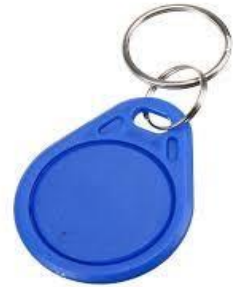

*Figure 2: RFID cards*  *Figure 3: RFID tag*

In conclusion, the purpose of this thorough literature research is to develop a thorough knowledge of the RFID-based smart attendance system. The review provides a thorough discussion of the issues and the solutions to them through a thorough investigation of the literature. For organizations wishing to use this cutting-edge technology, the thorough analysis of guiding principles, best practices, and implementation challenges pertaining to IoT-based attendance systems provides insightful information. This system is a major improvement in attendance management due to its real-time attendance tracking, smooth data integration, and increased accuracy. Educational institutions may promote student engagement, improve academic achievements, and expedite administrative procedures by embracing this cutting-edge technology. A more effective and efficient method of tracking attendance in educational settings is one of the many potential advantages of the IoT-based attendance system.

## 1.1. Background:

A system with hardware nodes based on RFID technology was illustrated by Zhang Yuru, Chen Delong, and Tan Lipping in 2013. The system's design greatly increases production and decreases resource waste, both material and human. The enormous potential of RFID-based technologies has already been recognized and addressed in previous publications. When a student reaches the campus, an SMS may be sent to a guardian using the GSM application that has been researched (Joshi et al., 2021). With the use of a central database and finger prints, Farzana Akter and her team proposed using the Internet of Things (IoT) to monitor student attendance (Mon et al., 2019). Only recently has the importance of RFID technology for controlling commercial supply chains been realized (Kurniali, 2014). Barcodes are among the most well-known and well-used systems now in use. It is an automated scanning system that looks like parallel lines of various widths and saves data about an object (Choudhury, 2017).

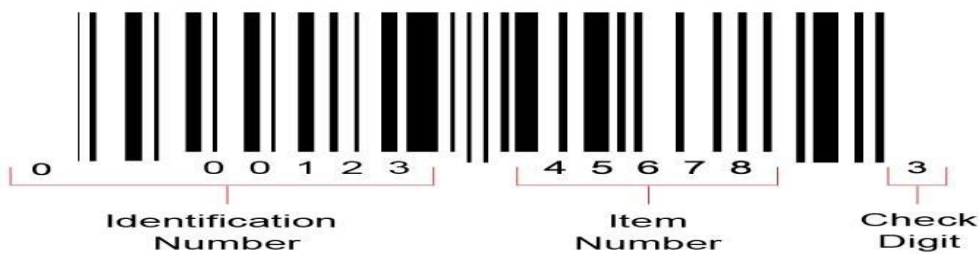

*Figure 4: Barcode*

This technology has a flaw in that barcodes may be readily copied by printed using an ordinary printer, despite the fact that it is quick enough and inexpensive (Dedy Irawan et al., 2018). The barcode reader is the most extensively used and well-liked existing system. It is made up of parallel, varying-width lines that function as an automated scanning system to record specific information about items. The item is tracked and identified using the electronic information contained in these parallel lines. A barcode card is a surface that has a barcode printed on it; they may be found on a variety of items, including digital gadgets, papers, plastics, and even textiles, such as shampoo bottles.

The Internet of Things (IoT) is a new paradigm in which everything may be connected to other things in order to immediately identify physical items and transport, store, and analyze information about them (El Mrabet & Moussa, 2020). The reader and the tags are the two essential elements of an RFID system. The tag often has a microchip that contains information and is affixed to the monitored things. On the other hand, the reader finds tags that fall within its detection range and either writes to or reads from them (Shah & Abuzneid, 2019). To conclude and authenticate students and workers of an organization, attendance records are required. As a result, several studies have been conducted in this field to enhance and replace the current system of attendance with RFID technology (Shah & Abuzneid, 2019).

**Table 1: Research questions**

| | RQ Statement | Objectives and Motivation |
|---|---|---|
| RQ1 | Which regions have been targeted by Attendance system using RFID research throughout the years, and what are the top publications channels for Attendance system using RFID? | The goal of RQ1 is to find high-quality research papers for the RFID attendance system using the main publications channels, taking into account geographical areas and publication trends over time. |
| RQ2 | What is the quality rating of the pertinent articles, target Attendance system employing RFID, its procedures, and research challenges? | Utilizable data on the framework/model, approach, and results were gleaned from the quality evaluation of the chosen papers and the meta-information. |
| RQ3 | What are the key components, features and benefits of implementing smart attendance system, and how does it contribute to creating sustainable and efficient attendance system? | You can monitor staff hours with the assistance of a smart attendance management system, ensuring that no rules or regulations are violated. An effective and reliable attendance system helps in managing absences and keeping track of employees' timeliness. |
| RQ4: | How can the integration of various IoT devices and technologies improve the quality of attendance for schools, colleges, and institutions with smart attendance systems? | IoT has several benefits for smart systems. IoT enables institutions to collect and analyze enormous volumes of data by integrating numerous devices, sensors, and systems, resulting in more effective operations and improved services for the attendance system. |
| RQ5: | What impact does the incorporation of smart system solutions have on the overall growth and effectiveness of the smart attendance system? | Smart systems provide multiple advantages that improve the presence quality of attendance by combining various IoT devices. As an illustration, consider connectivity, time consumption, energy management, safety, and security, ease of use, and efficient. |
| RQ6: | What are the main guiding principles, best practices, and difficulties discovered in IoT in the attendance system? | To recognize the underlying principles and procedures that can ensure the successful and efficient implementation of IoT in smart attendance systems. Additionally, there are a number of issues that must be resolved for the implementation to be successful, including interoperability, scalability, data security, and privacy, cost and funding, the absence of manual systems, and administrative and juridical structures. |

*Table 2: Method of searching*

| Electronic library | Search query | Used filter |
|---|---|---|
| **Web of Science (Core Collection)** | TITLE: (attendance system) OR TITLE: (smart attendance system) OR TITLE: (student attendance system) OR TITLE: (RFID based attendance system) OR TITLE: (RFID) OR TITLE: (IOT based attendance system) OR TITLE: (internet of things) OR TITLE: (smart attendance system rfid based) AND TITLE: (Comprehension) OR TITLE: (smart system) AND TITLE: (attendance) AND TITLE: (sensor) AND TITLE: "advance attendance system" OR "Institute" OR "University" OR "College" OR "School" Defining | **2015-2023** |

**Choosing according to inclusion/exclusion criteria**

**1. Selection standards**

The review's report must take into account students' attendance, absences, and presences while concentrating on the study themes. In addition, papers submitted at conferences or publications from 2015 to 2023 were taken into account. The evaluation also featured papers on smart attendance at the high school, college, and university levels that were concerned with evaluating, controlling, and the student's attendance.

**2. Exclusion Standard**

Articles that weren't in English or that didn't focus on using Smart Attendance systems in schools, colleges, and universities to teach and study English were omitted.

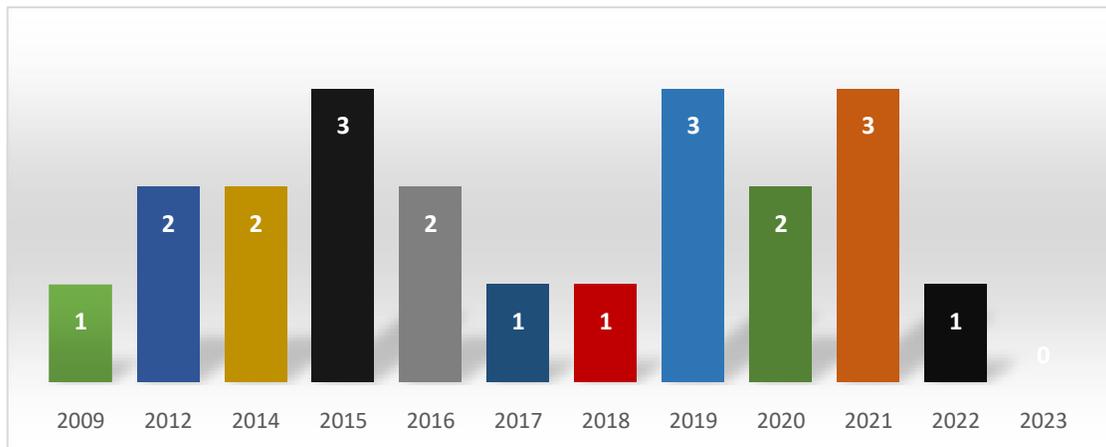

**Figure 5:** *Year-by-year graph of the selected studies*

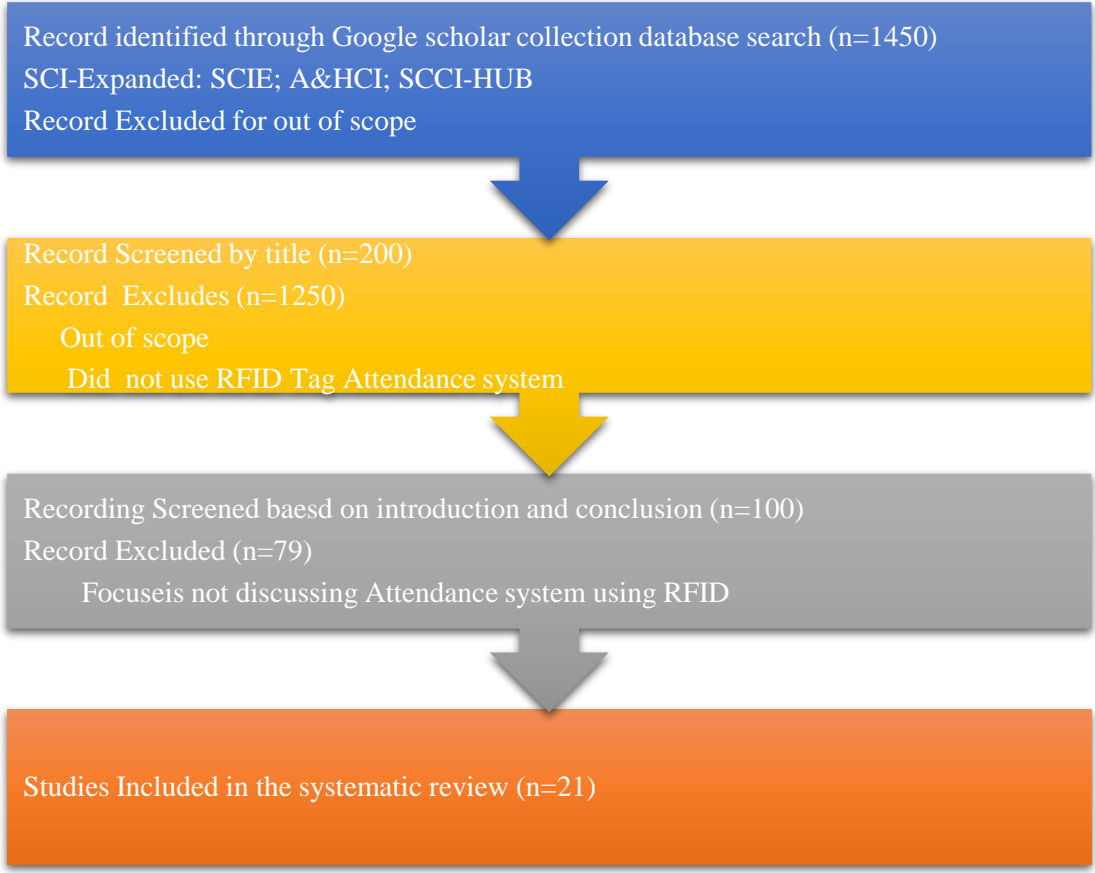

**Figure 6:** *Choosing pertinent articles with the help of the Systematic Review Process*

## 2. EVALUATION AND CONVERSATION OF RESEARCH QUESTIONS

This part examined 21 primary research papers that had been completed and were based on our research topics.

**Table 3:** *steps of selection and outcomes.*

| Sr. | Select | Selection standards | Index's:, Google Scholar, Research gate SCI-HUB, Peerj.com |
|---|---|---|---|
| 1 | Search | Keyword | 1450 |
| 2 | Filtering | Title | 200 |
| 3 | | Abstract | 150 |
| 4 | | Introduction and summary | 70 |
| 5 | Inspection | Full article | 21 |

**RQ1: Which regions have been targeted by Attendance system using RFID research throughout the years, and what are the top publications channels for Attendance system using RFID?**

For academics in the fields of education, technology, and industry, the integration of an attendance system into learning tools, techniques, material, and the choice of a theoretical viewpoint presented a considerable problem. It was important to find trustworthy publication platforms and do scientometric analysis based on meta-data inside the IoT domain in order to solve this difficulty. This portion emphasized developing in-depth information about the sources, categories, and years of research publications as well as how selected studies were dispersed throughout different publishing channels and among grade levels and regions. The goal of this analysis was to provide readers a thorough grasp of IoT research in the context of integrated smart systems and attendance. Each year, the research from the core collection of Web of Science were examined and presented in **Table 6** and **Figure 6.** The maximum number of publications, i.e. three, were chosen from the years 2015, 2019, and 2021 out of a total of 21 publications. The use of smart attendance systems in education, technology, and learning suggests a greater interest in the growth of IoT. However, there was less interest shown in the years 2009, 2017, 2018, 2022, and especially in 2023, which led to little advancement in improving teaching and learning in order to better meet the demands of students and the market.

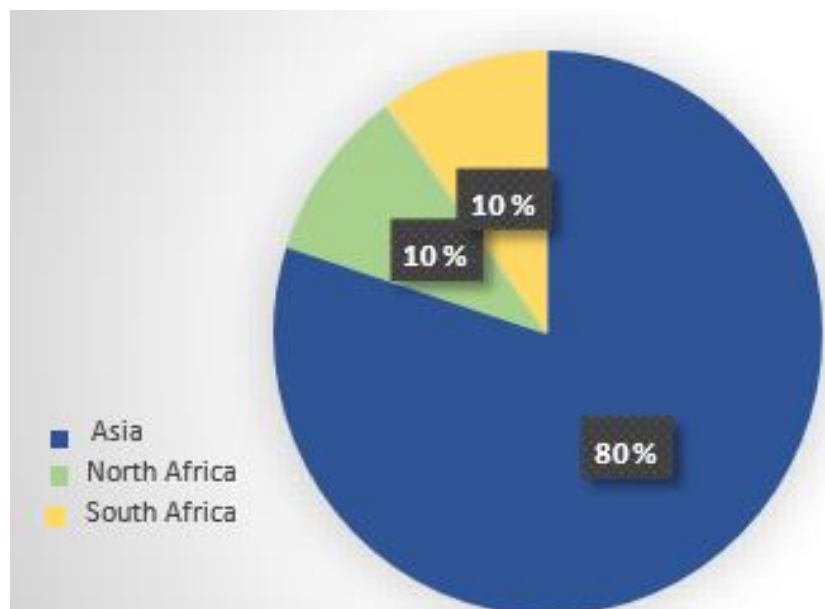

*Figure 7: Graph of the studies found, sorted by continent.*

*Table 4: Publications broken down by region*

| Items No. | Region | Nations | Publication number |
|---|---|---|---|
| 1 | Asia | Malaysia | 2 |
| | | Indonesia | 3 |
| | | India | 8 |
| | | Pakistan | 1 |
| | | USA | 2 |
| | | Bangladesh | 1 |
| | | China | 1 |
| | | Kazakhstan | 1 |
| 2 | North Africa | Morocco | 1 |
| | | South Africa | 1 |
| Total | | | 21 |

**Table 4** and **Figure 7** shows how research is distributed throughout various geographic areas. 19 of the 21 publications, or the bulk, were from different Asian nations. In addition, one research from Africa and one from North Africa were published. According to statistics in Table 5, the majority of papers were taken from famous publications that are listed in Google Scholar. One single piece was taken from a reputable seminar. With seven papers chosen, the Computer & Education journal came in first, followed by the Educational Technology & Society journal.

*Table 5: sources for publications*

| Item | Sources for publications | Channel | No. of articles |
|---|---|---|---|
| 1. | 2012 IEEE Control and System Graduate Research Colloquium (ICSGRC) | Conference | 1 |
| 2. | EPJ Web of Conferences | | 2 |
| 3. | 2009 IEEE Symposium on Industrial Electronics and Applications (ISIEA 2009) | | 1 |
| 4. | 2019 3rd International Conference on Energy Conservation and Efficiency (ICECE) | | 1 |
| 5. | International Conference on Advanced Computing & Communication Systems (ICACCS) | | 1 |
| 6. | IEEE – 40222 | | 1 |
| 7. | International Conference on electronics and communication system (ICECS 2015) | | 1 |
| 8. | Online international conference on green engineering and technologies (IC-GET) | | 1 |
| 9. | IJCSI International Journal of Computer Science Issues | | 1 |
| 10. | International Conference on electronics and communication system (ICECS 2015) | | 1 |
| 11. | Online international conference on green engineering and technologies (IC-GET) | | 1 |
| 12. | IJCSI International Journal of Computer Science Issues | | 1 |
| 13. | Journal of Physics | | 1 |
| 14. | MATEC Web of Conferences 164, 01020 (2018) | Journal | 1 |
| 15. | Computer Engineering | | 1 |
| 16. | International Association of Online Engineering | | 1 |
| 17. | International Journal of Computer Applications | | 1 |
| 18. | International Journal of Smart Home | | 1 |
| 19. | International Journal of Engineering Research & Technology (IJERT) | | 1 |

**Q2:** What is the quality assessment of the relevant publications, target Attendance system using RFID, its practices and challenges research?

Each completed study received a QA score in accordance with the guidelines in "Research Methodology" section B.3, as shown in Table 6. Less than four QA ratings were disregarded, with values ranging from 4 to 8. This QA score may be helpful to IoT researchers in choosing pertinent studies and resolving their usage and difficulties. The majority of articles published in Q1 journals obtained the highest score, while a few lesser-known but nonetheless pertinent publications received a score of 4. 19 of the 21 experiments received the top possible score of 8, demonstrating compliance with all QA requirements. Nine studies, however, were given a score of 4, which is the set's lowest QA rating.

**Table 6: Rating for stable and recognized publication sources**

| Number | Publication of Source | +4 | +3 | +2 | +1 | 0 |
|---|---|---|---|---|---|---|
| 1 | Journal | Q1 | Q2 | Q3 | Q4 | Not JCR Ranking |
| 2 | Conference | Core A | Core A | Core B | Core C | Not Core Ranking |

**Table 7** demonstrates the thorough categorization output and QA (Quality Assurance) of the completed research. The stated research kinds served as the foundation for the taxonomy that is described in section V. To raise the bar for quality, points were given to research that carefully confirmed their findings by statistical analysis, experiments, surveys, or case studies. Only 8 out of the 21 evaluations in category (c) of the quality assessment criteria did not provide empirical results, earning them a score of 0. Similar to category (d) of the quality evaluation criteria, only five studies received a score of 0, while the others received higher values, demonstrating their suitability as trustworthy sources. The four studies with the lowest scores each earned one.

RFID is increasingly being used in a variety of industries, including transportation, agriculture, and manufacturing. The laborious manual attendance system is one of the biggest issues in educational institutions. Our intention is to use RFID technology to resolve this issue. The essential components of the RFID system are RFID tags, readers, a backend storage system, and a central section that houses all the electrical components. When it was initially presented in the late 1960s, RFID technology wasn't particularly useful (Choudhury, 2017). We developed the questionnaire as an essential instrument to identify and comprehend the most significant obstacles to tracking attendance in Moroccan schools. At various school levels, the study focused on 411 teachers and superintendents. By analyzing the approaches to managing absenteeism in schools, this survey intends to investigate the causes of school absence. Our survey consists of a series of inquiries sent frequently to a sample of instructors in order to gather data on how to address student absence (El Mrabet & Moussa, 2020).

The students might be given RFID tags with their roll numbers on them. A copper coil within an RFID reader serves as an antenna. Due to the mutual inductance energy, data is conveyed to the reader when the tag is brought close to it. The microcontroller receives the data after which it continually checks for it. After then, the data is kept in a database. If the tag is legitimate, only then does the microcontroller take the attendance (Joshi et al., 2021). A number of repairable units make up each circuit's whole arrangement. The selection of materials was focused on budget, competency, robustness, and availability in order to have a project that was deserving, long-lasting, and cost-effective. This article explores the creation of an intra-connected configuration of RFID readers inside a learning environment using a typical structural design. Users may then readily access the registered data that is kept in the database and approach the data for further administration using applications such as Microsoft Excel (Bakht et al., 2019).

A label, which is actually a little silicon chip and radio wire, is used to identify anything. RFID scanners use radio waves to scan the labels on moving or stationary objects. The RFID reader must scan over the label on which the data is stored before sending it to a database, which interprets the label (Senthilkumar et al., 2020). The hardware's job is to read a signal produced by the tag in order to determine the tag ID. The database will receive the tag data before sending it on to the software program for additional data processing and analysis. The program is in charge of handling database-related tasks, processing data, and generating output in the form of reports and graphs. Numerous services might be employed once data has been gathered and saved in the database (Shah & Abuzneid, 2019).

## Table 7: Quality Control

| Ref. | Classification | | | | | Quality Control | | | | |
|---|---|---|---|---|---|---|---|---|---|---|
| | P. Channel | Year Publication | Type of Research | Method | Methodologies | (a) | (b) | (c) | (d) | Score |
| 1. | Research Journal | 2009 | Solution Suggestions | Experiment | Prototype | 1 | 0 | 1 | 1 | 3 |
| 2. | | 2012 | Research Evaluation | Experiment and Survey | Framework proposed | 1 | 2 | 1 | 4 | 8 |
| 3. | | 2012 | | Survey | | 1 | 1 | 1 | 2 | 5 |
| 4. | | 2014 | | Experiment | | 1 | 2 | 1 | 2 | 6 |
| 5. | | 2014 | | Survey | | 1 | 2 | 1 | 4 | 8 |
| 6. | | 2015 | | Experiment | Learning tool | 1 | 2 | 1 | 4 | 8 |
| 7. | | 2015 | | Survey | | 1 | 2 | 1 | 4 | 8 |
| 8. | | 2015 | Solution Suggestions | Experiments | | 1 | 1 | 1 | 3 | 6 |
| 9. | | 2016 | Research Evaluation | | | 1 | 1 | 1 | 1 | 4 |
| 10. | | 2016 | Solution Suggestions | | | 1 | 2 | 1 | 4 | 8 |
| 11. | | 2017 | Research Evaluation | Experiment + Survey | | 1 | 2 | 1 | 4 | 8 |
| 12. | | 2018 | Solution Suggestions | Experiment + Survey | Framework proposed | 1 | 1 | 0 | 1 | 3 |
| 13. | | 2019 | | Experiment | Methodology | 1 | 2 | 1 | 4 | 8 |
| 14. | | 2019 | | Survey | Learning tool | 1 | 2 | 1 | 4 | 8 |
| 15. | | 2019 | Solution Suggestions | Experiment and Survey | Framework proposed | 1 | 1 | 1 | 1 | 4 |
| 16. | | 2020 | | Experiment and survey | | 1 | 4 | 1 | 2 | 8 |
| 17. | | 2020 | | Framework | | 1 | 0 | 0 | 4 | 5 |
| 18. | | 2021 | Research Evaluation | Survey | Methodology | 2 | 1 | 1 | 0 | 4 |
| 19. | | 2021 | | | | 2 | 1 | 2 | 1 | 6 |
| 20. | | 2021 | | | | 1 | 1 | 1 | 1 | 4 |
| 21. | | 2022 | | | | 1 | 2 | 1 | 4 | 8 |

## Table 8. Summary

| Sr. No | References | Score |
|---|---|---|
| 1 | (Choudhury, 2017; Nurbek & Selim, 2012; Olanipekun & Boyinbode, 2015; Santoso & Sari, 2019; Senthilkumar et al., 2020; Shah & Abuzneid, 2019; Sharma & Aarthy, 2016; Singh et al., 2015; Singh et al., 2022; Yadav & Nainan, 2014) | 8 |
| 3 | (Makutunowicz & Konorski, 2014; Mishra et al., 2015; Naen et al., 2021) | 6 |
| 4 | (Kassim et al., 2012; Revathi et al., 2020) | 5 |
| 5 | (Santoso & Sari, 2019; Sharma & Aarthy, 2016; Ula et al., 2021) | 4 |

**RQ3: What are the key components, features and benefits of implementing smart attendance system, and how does it contribute to creating sustainable and efficient attendance system?**

The **Table 8** offers a comprehensive overview of the key components, features, and benefits entailed in deploying a smart attendance system. This sophisticated system harnesses diverse hardware and software elements to establish a streamlined and eco-friendly attendance recording process in educational institutes.

*Table 8: Benefits and Features*

| Aspect | Key Components | Features | Benefits | Contribution to Sustainable and Efficient Attendance System |
|---|---|---|---|---|
| Hardware Components | RFID Readers, Biometric Scanners, etc. | Automated Data Capture | Accurate and Real-time Attendance Recording | Reduces Paper Usage and Waste |
| | | Touch less Authentication | Enhanced Security and Prevention of Fraud | Lower Environmental Impact |
| | | High Speed Data Transfer | Swift and Efficient Data Processing | |
| Software Functionality | Cloud-based Database, Analytics Tools | Cloud Storage and Management | Centralized Data Access and Storage | Minimizes Physical Storage Requirements |
| | | Data Analytics and Reporting | Advanced Attendance Analysis and Insights | Facilitates Informed Decision Making |
| | | Mobile Application for Users | Convenient Access for Students and Faculty | Reduces the Need for Printed Attendance Lists |
| Integration and Scalability | API, Integration with LMS | Integration with Existing Systems | Seamless Incorporation with Educational Infrastructure | Maximize the Use of Existing Recourses |
| | | Scalability and Expandability | Easily Accommodates Growing Institutes and Campuses | Future Proof System with Room for Expansion |
| Environmental Impact | Energy Efficient Hardware | Low Power Consumption | Reduced Energy Usage and Carbon Footprint | Contribution to Sustainability Goals of the Institution |
| | | Paperless Attendance Processing | Minimizes Paper Waste and Environmental Impact | Supports Eco-friendly Initiatives of the Institution |

| Data Privacy and Security | Encryption Data Transmission | Data Protection Measures | Ensures Confidentiality and Privacy of Attendance Data | Mitigates Risks of Data Breaches and Unauthorized Access |
|---|---|---|---|---|
| | | Role-based Access Control | Restricts Unauthorized Data Access | |
| | | Secure Authentication Methods | Ensures Only Authorized Users Can Marks Attendance | |

Incorporating a smart attendance system yields a multitude of advantages, encompassing precise attendance records, heightened security, effortless data accessibility, and well-informed decision-making. Furthermore, the system's sustainable attributes, such as diminished paper usage and minimized energy consumption, harmonize with the institution's environmental objectives, rendering it a valuable asset in crafting an efficient, eco-friendly, and progressive attendance system. In most colleges, instructors take attendance by calling out students' names and surnames, marking them, whereas in other institutions, instructors dole out sheets of paper and instruct students to sign the sheet of paper next to their surnames. Both methods have disadvantages. In the first scenario, verifying all of the students by name and surname who come to the session in several groups may take around ten minutes out of each lesson; in the second scenario, friends of missing students can jot down the names and surnames of the students who aren't there. When it comes to recording attendance, these procedures significantly harm university instructors and their organizations. We have chosen to use the RFID-card in order to address these recurring issues. Each card has a unique ID that prevents card duplication (Nurbek & Selim, 2012).

The RFID scanner instantly reads a student's ID as soon as they walk into the class and sends information to the PC. This ID is transmitted to the computer, where the system will use it to compare the information with data in the database. The system then uses the given internet network to upload the statistics data for the current ID to the database server. Data about students who are present or not in a certain lecture, as well as student attendance time, will be stored in a database on the server. Professors may obtain database information for use in assessing attendance (Shah & Abuzneid, 2019). The hardware and software components make up the two primary sections of the creation of the RFID-based automated attendance system. The RFID reader, tags, and host computer make up the hardware component. The host system program, created using VB.net and integrated with a Microsoft Access database, is the software component. The system allows the administrator or lecturer to log in and use the application, which keeps a record of the ID, time, and date of each student who enters the lecture hall for class. Additionally, it may register new students by utilizing each tag's ID (Shah & Abuzneid, 2019).

The two types of tags utilized are active tags and passive tags. The amount of energy that each tag has stored within it determines how long it can last. Active tags have two power options: they may link to a powered infrastructure or draw power from the internal battery. The advantage of passive tags is that they often do not require batteries and don't require maintenance. The tags have a long lifespan and are small enough to fit within a useful sticky label. A passive tag consists of three components: an antenna, a semiconductor chip, and some kind of encapsulation. When a passive

RFID tag is brought close to an RFID tag reader, it powers up and simultaneously transfers data to the reader (Konatham et al., 2016). The tag antenna absorbs the energy and transmits the ID of the tag. A gear motor, a DC motor, generates high torque at a low speed. The door is connected to a motor shaft. The DC motor is connected to the microcontroller by a driver. GSM devices are used to transmit mobile voice and data services. GSM utilizes TDMA to transfer speech or data in different time slots. A GSM modem is attached to the microcontroller. The microcontroller uses UART, which is controlled by AT instructions that are transmitted from the microcontroller to the modem (Konatham et al., 2016).

The hardware and software are required to construct the RFID Based Student Attendance System with Notification to Parents Using GSM. This system includes a professor and student automated attendance mechanism. Both the lecturer and the student must use their RFID cards to record their attendance while entering the classroom. After processing through the microcontroller, this attendance will be sent to a central computer where it will be kept. Notification of a student's absence from class will be given to the student's parents (Yadav & Nainan, 2014). The proposed system shows an automated roll call that may identify pupils who are present or not. This information is sent to cloud storage in order to keep track of each student's attendance and attendance status. Cloud storage, administration, and user node are the three components of work flow. User modules include sensors, and management modules (classrooms) recognize user sensor data (Mon et al., 2019).

When a student is registered, the management module sends tag data to cloud storage and verifies the tag to see if the student is registered. Valid student information is saved to cloud storage via the network and an internet application interface so that it may be accessed whenever necessary to display the student's status in real time. Locally, the management module operated. It gets data from the user node and transmits it over the Wi-Fi module to the cloud storage. This framework's suggested use of cloud storage includes a system with both storage space and an application. Data is obtained and saved to cloud storage using it. Cloud storage aids in preserving all presence information on each person's attendance status (Mon et al., 2019). The adoption of a smart attendance system offers potential to raise intelligence levels and advance ICT abilities (El Mrabet & Moussa, 2020).

**RQ4: How can the integration of various IoT devices and technologies improve the quality of attendance for schools, colleges, and institutions with smart attendance systems?**

The advent of IoT technologies has heralded a new era of progress in educational institutes, revolutionizing both classroom experiences and safety protocols. The varied array of implementations observed in universities and colleges serves as a testament to the remarkable adaptability and versatility of IoT solutions within educational settings. Embracing these cutting-edge technologies empowers institutes to cultivate highly interactive, streamlined, and secure learning environments, benefiting both students and staff alike as shown in **Table 9**.

*Table 9: Usage of IoT Devices and Technologies in Educational Institutes*

| Sr. | Educational institute | IOT technologies in Use | Purpose of IoT implementation |
|---|---|---|---|
| 1. | University | RFID reader, LED, Buzzer, LCD, Network, microcontroller. | Enhance Class-room Experience |
| 2. | | RFID reader, Computer, Graphical user interface, database | Improve Security and Safety |
| 3. | | RFID Tag/key, RFID Reader, GSM Module, database | Improve Security and Safety |
| 4. | | RFID Card, RFID Reader, Camera, PC, Database | uniqueness, stability, permanency and easily taking |
| 5. | | RFID reader, microcontroller, database, GSM. | save time and Streamline Attendance Tracking |
| 6. | | RFID tags, RFID card reader, Cloud, Camera, | Improve Security and Safety |
| 7. | College | Microcontroller (RTC), (EEPROM), Keypad, LCD, ZigBee module. Raspberry Pi. | Enhance Classroom Experience and Improve Security and Safety |
| 8. | | RFID reader, microcontroller, Arduino Uno, LCD, cloud database | Enhance Classroom Experience |
| 9. | University | RFID reader, Camera, microcontroller, LCD, Database | Streamline Attendance Tracking |
| 10. | | RFID tags, RFID reader, Database | Enhance Classroom Experience |
| 11. | | RFID tag, RFID reader, microcontroller, pc, buzzer, LCD | Improve Security and Safety |
| 12. | | Smart card, RF reader, database, Arduino, LCD, RTC module, solenoid Lock | Streamline Attendance Tracking and Improve Security and Safety |

IoT is a new paradigm in which everything can be connected to other things in order to immediately identify physical items and transport, store, and analyze data between the real world and the virtual world at any time and from any location (El Mrabet & Moussa, 2020). One of the automated identifying technologies in recent years is RFID. In order to fully utilize this technology, there is much research and development being done in this field. In the upcoming years, a number of new applications and research areas will continue to emerge. Numerous applications, including inventory management, product tracking during manufacturing and assembly, parking lot access and control, bank locker security systems, automatic toll collection systems (ATCS), library management systems (LMS), attendance management systems, etc., have made extensive use of RFID systems (Yadav & Nainan, 2014).

The essential components of the RFID system are RFID tags, readers, a backend storage system, and a central section that houses all the electrical components. This RFID-based attendance system is particularly user-friendly for commercial usage and contains a storage system that stores the individual identification number of the student or employee (Choudhury, 2017). The goal is to streamline the high school attendance tracking process, look at methods to cut down on absences, and improve cooperation between parents and the school. The Internet of Things (IoT) is a worldwide network architecture that links numerous items to the Internet in order to share information and perform intelligent identification. Kevin Ashton originally used the phrase in 1999 (El Mrabet & Moussa, 2020).

The use of RFID tags allows school/college administration to monitor student mobility across the campus (Shah & Abuzneid, 2019). The primary purpose of an RFID project is to automatically track employee or student attendance. The use of new technology helps to improve technique and reduce human mistake. The goal of the suggested system is to identify an automated method for tracking university student attendance. The class list is used to complete the roll call system for students, and it takes time and energy to check on students' attendance, thus the suggested approach suggests that teaching will take up a lot of teacher time and resources. Therefore, it is very important to employ effective and current methods in order to avoid wasting time and energy. Thus, an automatic and universal attendance system might be implemented. In comparison to the conventional approach of looking at students, an automated system can provide greater routine and efficiency (Mon et al., 2019).

Data input is more timely and accurate thanks to the automatic ID and data gathering technology known as RFID. Microchip and radio frequency technology are combined in RFID to provide a smart system for the identification, monitoring, security, and inventory of things (Aravindhan et al., 2021). In order to maintain track of the number of pupils present in classrooms, institutions, or other settings, attendance is used. It is a crucial component of upholding order among employees in a business and providing high-quality instruction in schools and universities. If someone deviates from the established norms, correct action may be done (Ali et al., 2022; Singh et al., 2015). Automatic attendance tracking saves a professor time and improves parent-child communication. This takes time and bothers the lecturer and students alike (Konatham et al., 2016).

**RQ5: What impact does the incorporation of smart system solutions have on the overall growth and effectiveness of the smart attendance system?**

The integration of smart system solutions yields significant benefits for the overall growth and effectiveness of a smart attendance system. Consider the following impacts:

**Streamlined Automation and Enhanced Efficiency**: Smart system solutions automate various attendance tracking processes, such as data collection, analysis, and reporting. This reduces the need for manual effort and ensures accurate and timely attendance management (Asadov, 2023).

**Real-time Tracking:** Smart attendance systems equipped with smart system solutions enable real-time tracking of attendance data. This feature delivers instant updates and notifications about attendance status, facilitating prompt intervention and improved decision-making (Srinidhi & Roy, 2015).

**Improved Data Accuracy and Reliability:** By incorporating smart system solutions, the smart attendance system minimizes human errors and inconsistencies in attendance data. Automated data collection methods, including biometric recognition and RFID technology, guarantee precise and dependable attendance records (Rana, 2021).

**Enhanced Security Measures:** Smart system solutions bolster the security of the attendance system by implementing features such as biometric authentication or facial recognition. These measures effectively prevent buddy punching or fraudulent attendance practices, ensuring that only authorized individuals are accounted for (Makutunowicz & Konorski, 2014).

**Scalability and Flexibility:** Smart system solutions offer scalability, allowing the attendance system to adapt and cater to the evolving needs of an organization. These solutions can effortlessly handle varying attendance volumes, making them suitable for both small and large scale applications (Kuppusamy, 2019).

**Advanced Data Analysis and Insights:** The integration of smart system solutions empowers the smart attendance system with advanced data analysis capabilities. By analyzing attendance data, organizations can discern patterns, trends, and insights that optimize workforce management, identify attendance patterns, and improve overall productivity (Inamdar & Aswani).

**Time and Cost Savings:** Smart system solutions streamline attendance management processes, resulting in time and cost savings. The automation of data collection and analysis eliminates the need for manual entry and calculation, freeing up administrative staff to focus on other critical tasks (Bharathy et al., 2021).

**Seamless Integration with Other Systems**: Smart system solutions seamlessly integrate with other systems, such as payroll or HR systems. This seamless integration facilitates efficient data exchange, eliminates duplicate data entry, and improves overall data accuracy (Zorić et al., 2019).

**Improved Compliance Adherence:** Smart attendance systems equipped with smart system solutions assist organizations in adhering to regulatory and compliance requirements. These solutions generate detailed attendance reports, audit trails, and compliance documentation, ensuring compliance with labor laws and regulations (Dixon & Abuzneid, 2020).

**Remote and Mobile Access:** Smart system solutions often provide remote and mobile access to attendance data. This feature enables authorized personnel to conveniently access attendance information from anywhere, facilitating flexible work arrangements and enabling real-time monitoring and management (Singhal & Gujral, 2012).

**Customization and Adaptability:** Smart system solutions provide organizations with the ability to customize the smart attendance system to align with specific organizational needs. Administrators can configure attendance rules, generate customized reports, and adapt the system to evolving requirements, ensuring a tailored solution (Arbanowski et al., 2004).

**Enhanced Employee Experience:** Implementing a smart attendance system with smart system solutions elevates the overall employee experience. Employees benefit from streamlined attendance processes, reduced paperwork, and increased transparency, resulting in heightened satisfaction and engagement among the workforce (LinkedIn).

**Predictive Analytics:** Advanced smart system solutions leverage predictive analytics algorithms to forecast attendance patterns, identify potential issues, and optimize resource allocation. This proactive approach empowers organizations to effectively manage attendance-related challenges and make data-driven decisions to enhance operational efficiency (LinkedIn).

**Environmental Sustainability**: By replacing manual attendance tracking methods with smart system solutions, organizations actively contribute to environmental sustainability. The reduction in paper usage and energy consumption associated with traditional attendance systems significantly lowers the carbon footprint, aligning with eco-friendly practices (Revathi et al., 2020).

**Analytics-driven Insights:** Smart system solutions offer comprehensive analytics and insights into attendance patterns and trends. Through the analysis of attendance data, organizations gain valuable insights into employee behavior, identify areas for improvement, and make informed decisions to optimize workforce management (Mridha & Yousef, 2021).

**Resource Optimization:** With smart system solutions, organizations can optimize resources based on attendance data. By understanding attendance patterns and demand, resources can be allocated efficiently, ensuring the appropriate number of staff members are present at any given time to meet operational requirements (Naen et al., 2021).

**Proactive Attendance Management:** Smart system solutions enable organizations to proactively manage attendance-related issues. Automated notifications can be configured to alert managers or supervisors about attendance discrepancies, facilitating prompt action, addressing concerns, and ensuring compliance with attendance policies (Kamel et al., 2021).

**Seamless Integration with Existing Infrastructure:** Smart system solutions are meticulously designed to seamlessly integrate with existing infrastructure, such as access control systems or time clocks. This integration streamlines the attendance management process, creating a unified and efficient system while reducing complexities and improving overall performance (LinkedIn).

**Mobility and Flexibility:** Smart system solutions often provide mobile applications or web interfaces, empowering employees to conveniently mark their attendance using smartphones or tablets. This mobility and flexibility enhance employee convenience, enabling accurate attendance tracking even for remote or off-site workers (Sandhya et al., 2022).

**Improved Decision-Making:** By harnessing the insights provided by smart system solutions, organizations can make data-driven decisions regarding workforce planning, scheduling, and resource allocation. These informed decisions lead to improved operational efficiency and productivity, maximizing organizational outcomes (Zhao et al., 2022).

**Promoting Employee Accountability**: Smart attendance systems equipped with smart system solutions foster a culture of employee accountability. By maintaining accurate and transparent attendance records, employees are well aware that their attendance is under close monitoring. This heightened awareness acts as a deterrent against absenteeism, late arrivals, and other attendance-related issues (Barth, 1992).

**Ensuring Enhanced Data Security**: Smart system solutions prioritize data security through the implementation of robust encryption, access controls, and data backup mechanisms. These measures safeguard attendance data, ensuring its protection and confidentiality. By complying with privacy regulations, organizations mitigate the risk of data breaches and maintain the integrity of attendance information (Mishra et al., 2015).

**Facilitating Continuous System Improvement**: Smart system solutions often incorporate feedback mechanisms and analytics tools, enabling organizations to gather insights on system performance and user experience. This valuable feedback serves as a basis for ongoing improvements to the smart attendance system, addressing usability or functionality gaps and ensuring continuous enhancements (Arulogun et al., 2013).

**Scalability to Accommodate Growing Organizations:** Smart system solutions offer scalability, making them suitable for growing organizations. As the workforce expands, the smart attendance system seamlessly scales to handle the increased volume of attendance data. This scalability ensures consistent system performance and accommodates the evolving needs of the organization (Samaddar et al., 2023).

**RQ6: What are the main guiding principles, best practices, and difficulties discovered in IoT in the attendance system?**

In the realm of IoT within the attendance system, a range of guiding principles, best practices, and challenges have emerged.

# Guiding Principles:

The guiding principles that govern the application of IoT in attendance systems center on fundamental values and essential considerations. These principles are designed to facilitate the seamless integration and operation of IoT Technologies in attendance tracking, promoting efficiency, security, and reliability. The following are some several key guiding principles to consider:

*Security and Privacy:* It is imperative to prioritize security measures and safeguard the privacy of attendance data. Robust encryption, access controls, and data anonymization should be implemented to ensure the integrity and confidentiality of data (Nguyen et al., 2022).

*Interoperability:* Emphasizing interoperability among diverse IoT devices and platforms enables seamless integration and efficient data exchange. Adhering to standardized protocols and open APIs fosters compatibility and facilitates system scalability (Abounassar et al., 2022; LinkedIn).

*Data Accuracy and Reliability:* Ensuring the accuracy and reliability of attendance data is crucial. Implementing redundant data collection mechanisms, such as employing multiple sensors or biometric authentication, helps minimize errors and inconsistencies (LinkedIn).

*Data Governance:* Implementing comprehensive data governance practices ensures the responsible and ethical management of attendance data throughout its lifecycle, including collection, storage, usage, and disposal (Rana, 2021).

*Scalability and Flexibility:* Designing the attendance system with scalability and flexibility as key considerations allows for seamless expansion and adaptation to changing organizational requirements. This encompasses accommodating new IoT devices, managing increased data volumes, and integrating with other systems (Kuppusamy, 2019).

*Standards and Interoperability:* Adhering to industry standards and promoting interoperability among IoT devices and systems ensures compatibility, seamless integration, and future scalability (LinkedIn).

*User-Centric Design:* Placing emphasis on user experience and designing the attendance system with the end-users in mind fosters user adoption and engagement. Intuitive interfaces, clear instructions, and simplified workflows enhance usability (Nguyen et al., 2022).

# Best Practices:

The phrase "best practices in IoT for the attendance system" refers to a group of suggested methods and tactics that are intended to maximize the use and use of IoT technology for tracking attendance. These procedures have been carefully designed to raise the IoT-based attendance system's overall efficacy, efficiency, and performance. Here are a few essential best practices to think about:

*Scalable Architecture:* Designing a scalable architecture allows the attendance system to accommodate growing data volumes as the organization expands. Leveraging cloud-based solutions and edge computing technologies aids in managing large-scale deployments effectively (Nguyen et al., 2022).

***Real-time Data Processing:*** Processing attendance data in real time enables instant updates and notifications. Employing edge computing or edge analytics minimizes latency, ensuring prompt system responsiveness (Srinidhi & Roy, 2015).

***Data Analytics and Insights:*** Leveraging data analytics techniques allows organizations to derive valuable insights from attendance data. Identifying attendance patterns, optimizing resource allocation, and detecting anomalies enhance decision-making and operational efficiency (Mridha & Yousef, 2021).

***User-friendly Interfaces:*** Developing user-friendly interfaces for both administrators and employees simplifies system usage and encourages widespread adoption. Intuitive dash boards, mobile applications, and self-service options enhance the user experience and drive engagement (Nguyen et al., 2022).

***Energy Efficiency:*** Optimizing the energy consumption of IoT devices and infrastructure not only reduces operational costs but also supports sustainable practices. By incorporating power saving features, efficient communication protocols, and effective device management strategies, organizations can enhance energy efficiency (Naen et al., 2021).

***Remote Monitoring and Maintenance:*** Leveraging remote monitoring and maintenance capabilities minimizes the need for physical intervention. Remote diagnostics, firmware updates, and proactive device management enhance system uptime while minimizing disruptions (Singhal & Gujral, 2012).

***Data Analytics for Decision-making:*** Employing advanced data analytics techniques such as machine learning and predictive analytics on attendance data empowers organizations to extract meaningful insights. These insights enable informed decision-making in areas such as workforce management, resource allocation, and operational planning (Nguyen et al., 2022).

***Redundancy and Resilience:*** Implementing redundant components and backup systems ensures high availability and resilience. This includes redundancy in data storage, failover mechanisms, and the establishment of comprehensive disaster recovery plans to mitigate service disruptions (Nguyen et al., 2022).

***Regular System Maintenance:*** Implementing routine maintenance procedures, including device updates, security patches, and performance optimizations, ensures the ongoing reliability and stability of the IoT-based attendance system (Nguyen et al., 2022).

***Data Backup and Recovery:*** Implementing regular data backup mechanisms and disaster recovery plans safeguards attendance data from potential loss or corruption. This includes offsite backups, redundant storage systems, and data replication strategies (Nguyen et al., 2022).

## Difficulties:

In the context of IoT in the attendance system, difficulties encompass the array of challenges and obstacles that organizations might confront while deploying, running, and maintaining IoT technologies for attendance tracking. The nature of these difficulties may fluctuate based on the unique circumstances and intricacies of the IoT-based attendance system. Among the common challenges encountered are the following:

***Connectivity and Network Reliability:*** Ensuring reliable connectivity between IoT devices and the network infrastructure can be challenging, especially in remote or complex environments. Implementing network redundancies and robust connectivity solutions becomes crucial for maintaining seamless system functionality (Nadhan et al., 2022).

***Data Security and Privacy Risks:*** IoT devices and data transmission channels introduce potential security vulnerabilities. Protecting data from unauthorized access, mitigating cyber threats, and complying with privacy regulations pose ongoing challenges (Abounassar et al., 2022).

***Data Integration and Interoperability:*** Integrating data from diverse IoT devices and platforms can be complex. Varied data formats, protocols, and compatibility issues necessitate meticulous planning and implementation of data integration strategies (Abounassar et al., 2022; LinkedIn).

***System Complexity and Scalability:*** Deploying and managing a large-scale IoT attendance system requires careful handling due to inherent intricacies. Ensuring system scalability, addressing infrastructure requirements, and effectively managing device provisioning and firmware updates demand comprehensive planning and allocation of resources (Abounassar et al., 2022; LinkedIn)..

***Change Management and Adoption:*** Implementing IoT technologies in the attendance system often necessitates organizational and cultural changes. Overcoming resistance to change, providing adequate training, and fostering user acceptance become vital for successful adoption (Sneesl et al., 2022).

***Interoperability Challenges:*** Integrating diverse IoT devices from different manufacturers and ensuring seamless interoperability can be a challenge due to varying protocols, standards, and compatibility issues. Overcoming these challenges requires meticulous planning and collaborative efforts (Abounassar et al., 2022; LinkedIn).

***Data Privacy and Consent:*** Managing data privacy and obtaining user consent can be complex, particularly when biometric data is involved. Organizations must navigate legal and ethical considerations, adhere to privacy regulations, and establish transparent policies regarding data collection and usage (Nguyen et al., 2022).

***Data Integration and Management:*** Integrating attendance data from multiple sources, such as Access control systems, biometric devices, and sensors, can be intricate. Ensuring data consistency, synchronization, and quality across different systems and platforms poses significant data management challenges (Zorić et al., 2019).

***Cost Considerations:*** The implementation of IoT-based attendance systems entails substantial upfront costs, including device procurement, infrastructure setup, and system integration. Organizations must conduct a thorough assessment of the return on investment (ROI) and carefully evaluate long-term cost implications (Santoso & Sari, 2019).

***Connectivity and Network Infrastructure:*** Establishing reliable connectivity between IoT devices and the network infrastructure can be challenging, particularly in large or geographically dispersed environments. Overcoming connectivity issues and ensuring adequate network coverage are critical (Hossain et al., 2019).

***Data Integration and Management Complexity:*** Managing and integrating attendance data from multiple sources, such as biometric devices, access control systems, and IoT sensors, can be complex. Organizations must address data format inconsistencies, data quality assurance, and data governance challenges (Hossain et al., 2019).

***System Security Vulnerabilities:*** IoT-based attendance systems may be vulnerable to security breaches, including unauthorized access, data manipulation, or denial-of-service attacks. Implementing robust security measures, such as authentication protocols, encryption, and intrusion detection systems, is essential (Hossain et al., 2019).

***Privacy and Compliance:*** Collecting and managing personal attendance data raises privacy concerns and requires compliance with applicable data protection regulations. Organizations must ensure compliance with privacy laws, obtain necessary consents, and establish transparent data handling practices (Hossain et al., 2019).

*Data Volume and Analytics:* Large amounts of attendance data created by IoT devices can be difficult to manage and analyze in terms of data storage, processing power, and gaining useful insights. To extract value from the data, organizations need use scalable data storage systems and cutting-edge analytics methods (Hossain et al., 2019).

## Taxonomy

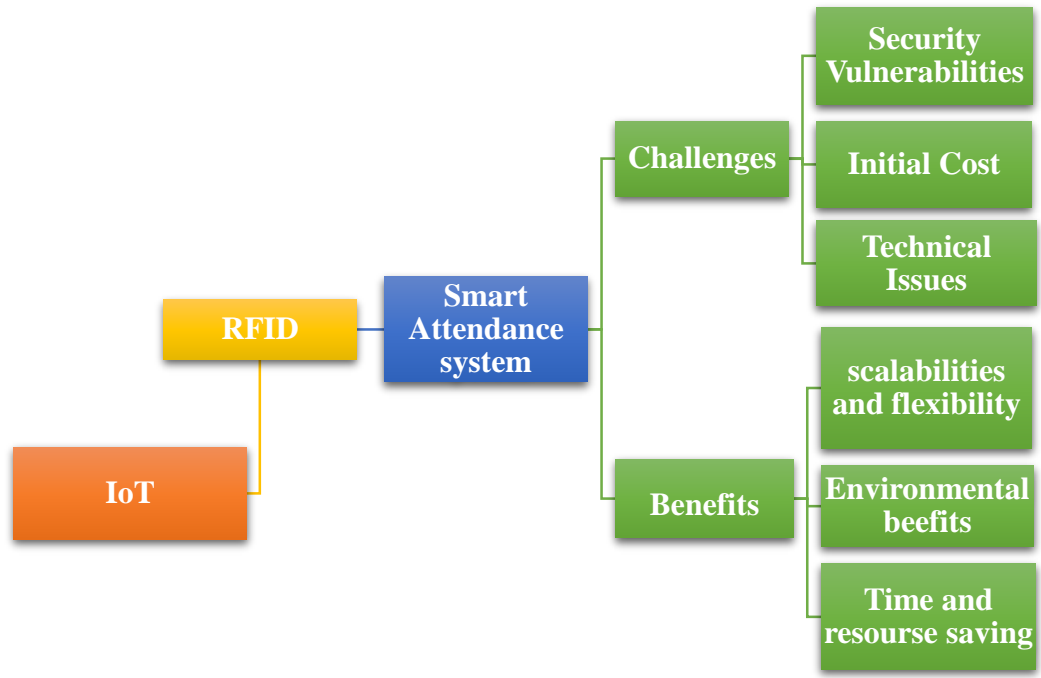

*Figure 8: Taxonomy*

The **Figure 8** depicts the intricate interconnections among diverse concepts in the realm of IoT, centering on RFID technology and its implementation in Smart Attendance Systems. Additionally, the diagram explores the challenges and advantages entailed in the adoption of these systems. The diagram highlights the substantial role of IoT, particularly with the integration of RFID technology, in advancing and implementing Smart Attendance Systems within educational institutions. Despite the existence of challenges, the benefits derived from these systems surpass any obstacles, providing educational institutions with an effective, secure, and technologically sophisticated method for managing attendance data.

## Conclusion:

This comprehensive study of the literature intended to discover emerging research directions in RFID-based IoT-based attendance systems. To provide a complete discussion of the problems and potential solutions in this field, the literature was carefully examined. The search was conducted using a wide range of well-known terms associated with smart attendance systems, and the outcomes were carefully reviewed. The search, which included research up to that point, was wrapped up in August 2023. In the investigation, 21 out of 1450 papers from the Google Scholar core collection were chosen for evaluation. Notably, all but one of the selected pieces came from conferences, and the bulk were published in respected publications. Most of the selected research was evidence-based, highlighting the numerous benefits that a Smart attendance system can offer for both teachers and students. Attendance system emerged as a frequent key aspect, with a particular emphasis on language learning strategies and evaluating students' performance. On the other hand, specific aspects of the attendance system emerged as a frequent key aspect, with a particular emphasis on language learning strategies and evaluating students' performance. The SLR found some shortcomings, especially in terms of study methods, the precision of data collecting, and probable misclassification. The method of employing distinct terms from the Google Scholar core collection repository, however, helped to reduce the possibility of selection mistakes. Specific inclusion/exclusion rules were adopted, and all data extractions were evaluated by two independent specialists in order to allay external concerns. This comprehensive evaluation of the literature offers important new information about the current state of study on RFID-based IoT-based attendance systems. The results highlight how these tools might completely change how attendance is tracked in educational settings. The use of an evidence-based methodology and an in-depth analysis of research trends expands the body of knowledge in this area and paves the way for smart attendance system enhancements in the future.